\documentclass[twocolumn]{aastex631}

\newcommand{\vsini}{$v\sin{i_*}$}

\newcommand{\teff}{\ensuremath{T_{\text{eff}}}}
\newcommand\kms{km~s$^{-1}$}

\newcommand{\gaia}{\textit{Gaia}}
\newcommand{\banyan}{\texttt{BANYAN $\Sigma$}}

\newcommand{\red}[1]{#1}

\begin{document}

\title{A Lithium Depletion Age for the Carina Association}

\correspondingauthor{Mackenna Wood}
\email{woodml@mit.edu}

\author[0000-0001-7336-7725]{Mackenna L. Wood}%
\affiliation{Department of Physics and Astronomy, The University of North Carolina at Chapel Hill, Chapel Hill, NC 27599, USA} 
\affiliation{MIT Kavli Institute for Astrophysics and Space Research Massachusetts Institute of Technology, Cambridge, MA 02139, USA}

\author[0000-0003-3654-1602]{Andrew W. Mann}%
\affiliation{Department of Physics and Astronomy, The University of North Carolina at Chapel Hill, Chapel Hill, NC 27599, USA} 

\author[0000-0002-8399-472X]{Madyson G. Barber}
\affiliation{Department of Physics and Astronomy, The University of North Carolina at Chapel Hill, Chapel Hill, NC 27599, USA} 

\author[0000-0002-9446-9250]{Jonathan L. Bush}%
\affiliation{Department of Physics and Astronomy, The University of North Carolina at Chapel Hill, Chapel Hill, NC 27599, USA}

\author[0000-0002-1312-3590]{Reilly P. Milburn}%
\affiliation{Department of Physics and Astronomy, The University of North Carolina at Chapel Hill, Chapel Hill, NC 27599, USA} 

\author[0000-0001-5729-6576]{Pa Chia Thao}%
\altaffiliation{NSF GRFP Fellow}
\affiliation{Department of Physics and Astronomy, The University of North Carolina at Chapel Hill, Chapel Hill, NC 27599, USA} 

\author[0000-0001-8510-7365]{Stephen P. Schmidt}%
\affiliation{Department of Physics and Astronomy, Johns Hopkins University, Baltimore, MD 21218, USA} 

\author[0000-0003-2053-0749]{Benjamin M. Tofflemire}
\altaffiliation{51 Pegasi b Fellow}
\affiliation{Department of Astronomy, The University of Texas at Austin, Austin, TX 78712, USA}

\author[0000-0001-9811-568X]{Adam L. Kraus}
\affiliation{Department of Astronomy, The University of Texas at Austin, Austin, TX 78712, USA}

\begin{abstract}

The dispersed remnants of stellar nurseries, stellar associations, provide unparalleled samples of coeval stars critical for studies of stellar and planetary formation and evolution. 
The Carina Stellar Association is one of the closest stellar associations to Earth, and yet measurements of its age have varied from $13$ to $45$ Myr. We aim to update the age of Carina using the Lithium Depletion Boundary (LDB) method. We obtain new measurements of the Li $6708$\AA\, absorption feature in likely members using optical spectra from the Goodman HTS on SOAR and NRES on LCO. We detect the depletion boundary at $M_K\simeq6.8$ (M5). This age is consistent within uncertainties across six different models, including those that account for magnetic fields and spots. We also estimate the age through analysis of the group's overall variability, and by comparing the association members' color-magnitude diagram to stellar evolutionary models using a Gaussian Mixture Model, recovering ages consistent with the LDB. Combining these age measures we obtain an age for the Carina association of $41^{+3}_{-5}$ Myr.
The resulting age agrees with the older end of previous age measurements and is consistent with the lithium depletion age for the neighboring Tucana-Horologium Moving Group.
\end{abstract}



\section{Introduction} \label{sec:intro}

Young stellar associations are valuable tools for studying stellar and planetary evolution. It is thought that all or nearly all stars form in associations with tens or hundreds of other stars as giant molecular clouds collapse, triggering star formation. These groups of coeval stars spread out over the first Gyr after their formation, dispersing until they are indistinguishable from the rest of the field. 
Prior to their full dispersal, young associations can serve as stellar laboratories, each providing a sample of stars with the same initial environmental conditions, age, and kinematic trajectory across a range of stellar types. Young coeval populations have facilitated the study of stellar evolutionary paths \citep[e.g.][]{bell_pre-main-sequence_2013}, protoplanetary disks and planet formation \citep[e.g.][]{fang_young_2013}, and young planet evolution \citep[e.g.][]{ciardi_k2-136_2018, bohn_young_2020, mann_tess_2022, bouma_kepler_2022}.

The Carina Association is one such young association. Discovered by \citet{torres_great_2001}, it was originally considered part of the Great Austral Young Association (GAYA) complex, which included two additional nearby associations, Tucana-Horologium and Columba. However, the relationships between these groups is uncertain. Tucana-Horologium may be independent from Carina-Columba \citep{kraus_stellar_2014}, or it may be part of the larger complex. \citet{gagne_number_2021} used expanded kinematic information from \gaia{} Data Release 3 (DR3) to search for connections between populations, finding that Carina and Columba are connected through the Theia 92 and 113 groups \citep{kounkel_close_2019} to Platais 8 \citep{platais_search_1998}. Additional evidence for a large complex in the region came from \citet{kerr_spyglass_2022}, which examined the connection between Carina, Columba, Tuc-Hor, Platais 8 and the Fornax-Horlogium association, finding that they can be linked as a single star-formation event with two ``cores".

The relationships between these groups are further complicated by uncertainty in their ages. The three groups originally thought to form the GAYA complex were estimated to be about 40 Myr old. Tucana-Horologium, the largest and best studied of the three, was confirmed to be $40$ Myr by \citet{kraus_stellar_2014}. However, the ages of Carina and Columba are less certain. 
The age of Carina has been measured several times using a variety of membership lists and methods, resulting in estimates of $13$ Myr \citep[isochrone, ][]{booth_age_2021}, $\sim 21$ Myr \citep[lithium sequence, ][]{schneider_acronym_2019}, $> 28$ Myr \citep[kinematic, ][]{miret-roig_dynamical_2018}, and $45^{+11}_{-7}$ Myr \citep[isochrone, ][]{bell_self-consistent_2015}. 

A tighter age constraint on the Carina association will help to determine its relationship with nearby associations, but it can also provide constraints on the formation and evolution of planets. Members of Carina host debris \citep{moor_new_2016} and protoplanetary disks \citep{silverberg_new_2016, murphy_wise_2018}, are dippers \citep{gaidos_planetesimals_2022}, and have been suggested as an origin of the interstellar object 'Oumuamua \citep{hsieh_evidence_2021}. At a mean distance of $75$ pc, Carina is also one of the closest known stellar associations, facilitating observational studies of such objects, including direct imaging and atmospheric characterization of planets.

A precise and fairly model-independent method for determining association age, the Lithium Depletion Boundary (LDB) method relies on locating the sharp cutoff between Li-rich and Li-poor late M dwarfs within the association. 
Lithium is depleted via proton-proton reactions in the cores of stars with core temperatures $>2.5\times10^6 K$.
As they approach the Main Sequence (MS), stars of different masses will reach this threshold temperature at different times. This results in a sharp boundary between those which have reached the threshold temperature and fully depleted their initial Li, and those slightly lower mass stars which have not and retain their full initial amount of Li. 
\red{In fully convective late-M objects, this depletion is rapid and complete, leading to a very sharp depletion boundary, the age-dependence of which is not strongly effected by model choice or initial Li abundance \citep{burke_theoretical_2004, tognelli_cumulative_2015}. By focusing on these very low mass objects, LDB is less model dependent than methods using the Li abundances of early-M or higher mass stars. In those partially-convective stars, the depletion boundary is not as well defined and depends on factors such as convective overshoot, such that uncertainty in measurements or model choice causes more age uncertainty.}

In this paper, we apply the Lithium Depletion Boundary (LDB) method, combined with an isochrone fit, empirical Li sequence comparison, and analysis of the group's variability, to measure the age of the Carina association.
In Section \ref{sec:membership}, we describe our selection of Carina members using \banyan{} and color-magnitude diagram (CMD) information. Our observational program is described in Section \ref{sec:observations}, and in Section \ref{sec:age}, we determine the age of the association. We conclude in Section \ref{sec:discussion} with a discussion and summary of our results.

\section{Membership Selection} \label{sec:membership}

To select the initial list of Carina members we use the \banyan{} Bayesian tool for determining membership probabilities. \banyan{} uses kinematic models of nearby young associations to calculate the probability of a given star's membership within each association or the field. We use the kinematic models defined by \citet{gagne_banyan_2018} for most of the associations, but due to the proximity of Carina to Lower Centaurus Crux (LCC), and MELANGE-4 \citep{wood_tess_2023}, we use an updated model for those groups. The new model of LCC, incorporating its sub-populations, and the model of MELANGE-4 are described in \citet{mann_tess_2022} and \citet{wood_tess_2023}, respectively. 

To construct the input sample for \banyan{}, we select from \gaia{} DR3 all stars within 100 pc of HD 49855, a high-probability member of Carina \citep{gagne_banyan_2018}. This sample comprised $542,642$ stars. We use this selected sample rather than the full \gaia{} DR3 200 pc sample to decrease the computation time and memory demands from the \banyan{} run, and to ensure that we included all stars near the Carina association. We use \gaia{} DR3 RA, Dec, parallax, proper motion, and, when available, radial velocity. 

From the results of \banyan{}, we take stars with membership probability $ P > 50\%$ and for which Carina is the best hypothesis (BESTHYP = CAR), yielding $129$ candidate members. 

\begin{figure}[tb]
    \centering
    \includegraphics[width=0.48\textwidth]{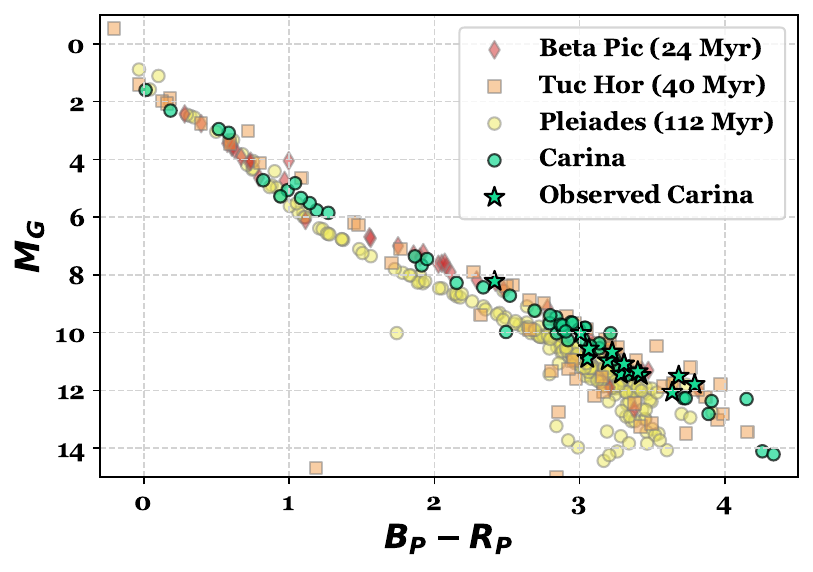}
    \caption{CMD of Carina members, overplotted with members of three benchmark associations, $\beta$ Pictoris ($24$ Myr), Tuc-Hor ($40$ Myr), and Pleiades (112 Myr). Membership lists for benchmark associations are generated using \banyan{} with the same parameters as described in Section \ref{sec:membership}. High probability members with $RUWE < 1.2$ are shown for benchmark associations. Observed members of Carina (see Table \ref{table:observed}) are indicated with stars.}
    \label{fig:carina_cmd}
\end{figure}

As \banyan{} only uses kinematic indicators of membership in determining the probability, the sample will include some older co-moving interlopers. To remove them we use the empirical definition of the MS given by \citet{pecaut_revised_2012}, removing candidates which have $B_P-R_P > 1$, $G - R_P > 0.5$, and are fainter than the MS (see Figure \ref{fig:carina_cmd}). 
These cuts leave a membership list of $99$ probable Carina members, shown in Figure \ref{fig:carina_cmd}.

\section{Observations} \label{sec:observations}
\subsection{SOAR/Goodman}
To obtain the medium-resolution optical spectra necessary for measuring the Li 6708 \AA{} line in low-mass association members, we use the Goodman High Throughput Spectrograph (HTS) on the 4.1m Southern Astrophysical Research Telescope (SOAR) located in Cerro Pachon, Chile. We observe a total of $15$ candidate members of Carina.
Observations were taken under mostly photometric conditions over five nights from 2020 Feb 6 through 2022 Feb 20.

These 15 stars were selected from the sample of candidate association members to map the LDB. We use an age estimate of $45$ Myr \citep{bell_self-consistent_2015} to predict the magnitude of the LDB. Comparing this age to BHAC 15 stellar evolutionary models \citep{baraffe_new_2015}, we calculate the expected 2MASS $K_S$ magnitude of $99\%$ Li depletion. Stars with $\Delta K_S < 1$ of this predicted boundary are selected for observation.
We update the observing list as needed as we obtain more data and revise the age estimate. Stars are prioritized for observation based on their \gaia{} RP magnitude and location on the sky. 
We omit potential binaries, as their inclusion could bias the measurement of age. We do so using the \gaia{} Renormalized Unit Weight Error (RUWE), a measure of the astrometric model goodness-of-fit. A high RUWE is indicative of binarity \citep{sullivan_undetected_2021, belokurov_unresolved_2020, wood_characterizing_2021}, so we remove those with $RUWE > 1.2$ from the observing list.

To measure the EW of the Li $6708$ \AA{} line, we use the Goodman red camera, 1200 l/mm grating, and the M5 mode, providing a wavelength coverage of $6300$--$7400$ \AA. We varied the slit width used between the $0.45\arcsec$ slit and the $0.6\arcsec$ slit depending on the target magnitude and the atmospheric seeing. This setup should give a resolving power of R = 4500–5800, although in practice the true resolution is lower and varies with exposure time (see below). For each target, we take five spectra with exposure times varying from 100 to 1600 s each.

To reduce the spectra we perform standard bias subtraction, flat-fielding, and optimal extraction of the target spectrum. 
Issues with the mount model and flexure compensation system cause large wavelength shifts during and between exposures, with shifts up to 5--10 pixels between subsequent exposures. Following \citet{wood_tess_2023}, we correct this effect using simultaneous skyline spectra and Ne arc spectra taken prior to each image. 
For a majority of targets we use the simultaneous skyline spectra to determine a fourth-order polynomial wavelength solution which is applied to calibrate each individual exposure. We then stack all exposures using a weighted mean. 
However, for targets with exposure times $t\lesssim300s$, the simultaneous skyline spectra do not have sufficient SNR to determine the wavelength solution, so we use a fourth-order polynomial wavelength solution derived from the Ne arc and corrected with a linear factor based on the simultaneous telluric absorption lines.
Each star is then corrected to its rest wavelength using radial-velocity standards taken with the same setup. While the resulting spectra were useful for determining spectral type and measuring EW(Li), we found that measured RVs had $\sigma_{\rm{rv}}\simeq5-10$ \kms{}, based on spectra of RV standard stars. Therefore, we do not use these spectra to measure RV for membership confirmation.  

\subsection{LCO/NRES}
To supplement our measurements of Li in low-mass association members we also obtain spectra of 3 higher-mass stars using the Network of Robotic Echelle Spectrographs (NRES) \citep{siverd_nres_2018} at the Las Cumbres Observatory. Observations were taken between 2022 Aug 22 and 2022 Sept 9.

The NRES spectra cover a wavelength range of 380--860 nm with high resolution ($R\sim 53,000$). The data are reduced using the LCO NRES pipeline \texttt{BANZAI-NRES}\footnote{\url{https://github.com/LCOGT/banzai-nres}}.

\subsection{Measuring EW} %
Radial velocities are extracted by cross-correlating the spectra with PHOENIX model atmospheres \citep{husser_new_2013} for stars of the same spectral class. 
After correcting the spectrum to the star's rest frame, we measure the EW of the Li 6708 \AA{} absorption line, used in Section \ref{sec:ew_li}. 

We estimate the pseudo-continuum by fitting a line to the spectrum in the region on either side of the Li feature. This pseudo-continuum does not account for molecular contamination of the continuum, or the nearby Fe absorption feature ($6707.4$\AA), which should have minimal contribution in M dwarfs.
The wavelength bounds of the line were adapted depending on the width of the line, which is effected by resolution differences between spectra and stellar \vsini. We did not assume a Gaussian profile for the absorption feature, instead performing a trapezoidal sum of the line underneath the estimated pseudo-continuum. Doing so allows us to account for any distortion from a Gaussian profile caused by the slightly different wavelength solutions between the added exposures.
Uncertainties were calculated by perturbing each point on the spectrum by its uncertainty (selected from a normal distribution) and re-calculating the EW(Li). This process was repeated 500 times, and the resulting median and standard deviation were taken as the EW(Li) and the uncertainty.
However, due to contamination of the continuum, the uncertainties for all targets are likely no better than 10\%. 

\begin{figure}[tb]
    \centering
    \includegraphics[width=0.48\textwidth]{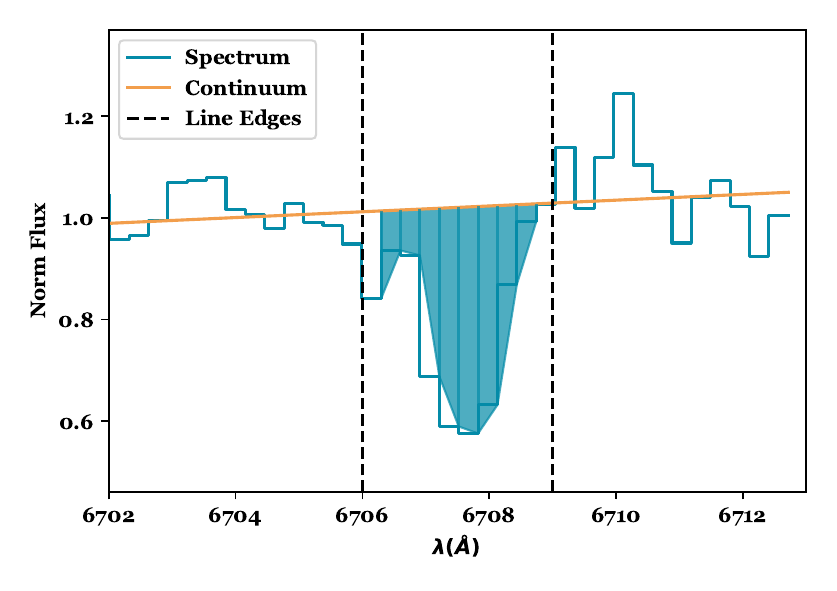}
    \caption{\red{Example Equivalent Width measurement. The pseudo-continuum, shown in orange, is a linear fit of the spectrum in the regions on either side of the line edges.
    The EW is calculated using a trapezoidal sum between the spectrum and the pseudo-continuum fit.}}
    \label{fig:eqw}
\end{figure}

\begin{figure}[]
    \centering
    \includegraphics[width=0.49\textwidth]{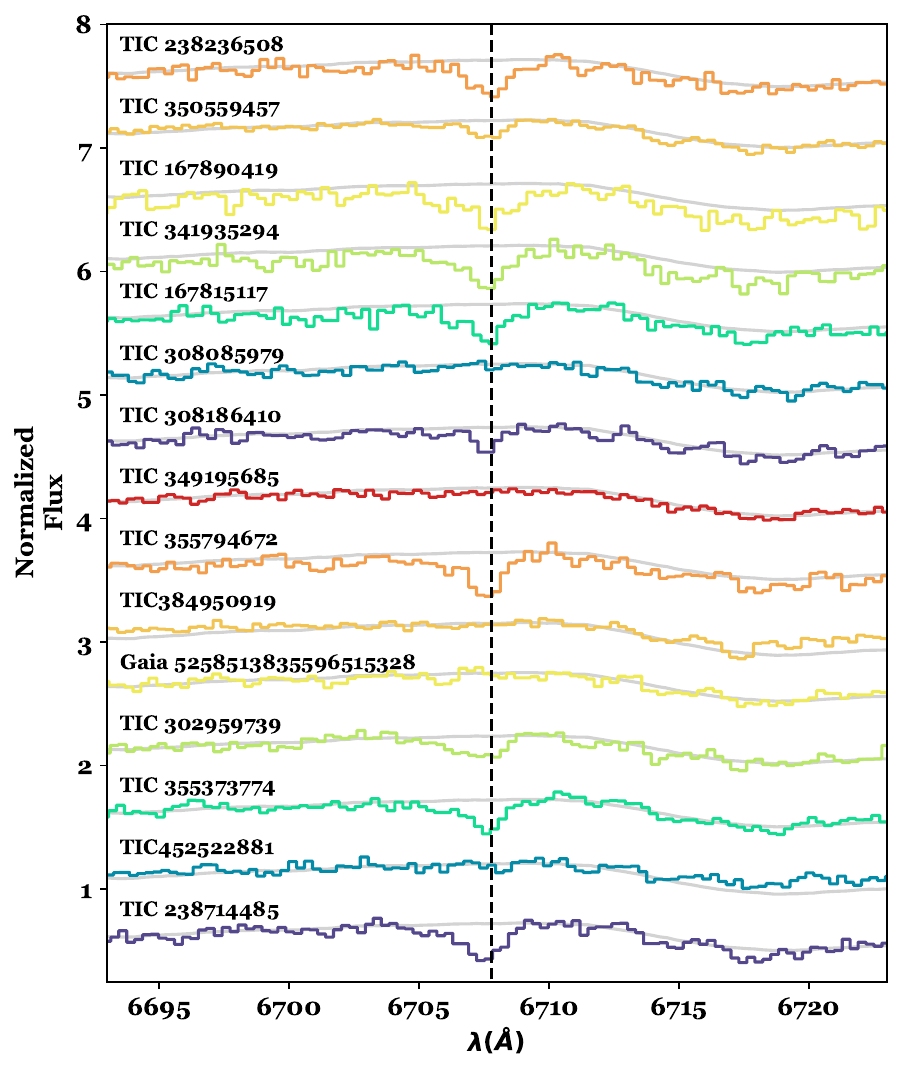}
    \caption{\red{Reduced SOAR/Goodman spectra of all observed Carina members, overplotted with a Li-free M-dwarf template (gray; \citet{bochanski_low-mass_2007}). The wavelength of the Li 6707.8\AA{} line is marked with a vertical dashed line. Four of these stars show either no detectable Li, or have EW(Li) $< 200$m\AA, and are thus Li-poor, while the other $11$ have  $EW(Li) > 200$m\AA, and are Li-rich.}}
    \label{fig:all_spectra}
\end{figure}

Observations and resulting Li measurements are listed in Table \ref{table:observed}. All SOAR/Goodman spectra are shown in Figure \ref{fig:all_spectra}.

\begin{table*}
    \centering
    \caption{Observations of Carina candidate members.}
    \vspace{5pt}
    \begin{tabular}{|llcccccc|}
    \hline
Object & Telescope & ObsDate & $T_{exp} $ & $M_{K_s}$ & $B_P - R_P$ & $G$ & EW(Li) \\
       &   &  YYYYMMDD & s & mag & mag & mag & m\AA \\
    \hline
TIC 238236508 & Goodman/SOAR & 20200206 & 1500, 1600 & 7.18 & 3.421 & 15.777 & $529.5 \pm 30.0$ \\
TIC 350559457 & Goodman/SOAR & 20200206 &        900 & 7.25 & 3.284 & 14.658 & $203.9 \pm 19.5$ \\ 
TIC 167890419 & Goodman/SOAR & 20200305 &        600 & 7.22 & 3.791 & 16.362 & $646.9 \pm 58.6$ \\ 
TIC 341935294 & Goodman/SOAR & 20200305 &        600 & 7.60 & 3.636 & 16.193 & $702.6 \pm 36.3$ \\
TIC 167815117 & Goodman/SOAR & 20220218 &        900 & 7.15 & 3.339 & 15.922 & $618.0 \pm 26.8$ \\
TIC 308085979 & Goodman/SOAR & 20220218 &        600 & 6.92 & 3.053 & 15.295 &            $< 10$\\
TIC 308186410 & Goodman/SOAR & 20220218 &   700, 600 & 6.87 & 3.193 & 15.620 & $226.4 \pm 25.2$ \\
TIC 349195685 & Goodman/SOAR & 20220218 &        600 & 6.60 & 3.064 & 15.307 &           $< 10$ \\
TIC 355794672 & Goodman/SOAR & 20220218 &        600 & 6.99 & 3.682 & 15.750 & $627.0 \pm 28.0$ \\
TIC 384950919 & Goodman/SOAR & 20220218 &   180, 120 & 4.73 & 2.415 & 13.184 &  $32.3 \pm 19.2$ \\
Gaia 5258513835596515328 & Goodman/SOAR & 20220218 & 200 & 6.05 & 3.007 & 13.811 &       $< 10$ \\
TIC 302959739 & Goodman/SOAR & 20220219 &        750 & 6.87 & 3.305 & 15.630 & $377.6 \pm 28.0$ \\
TIC 355373774 & Goodman/SOAR & 20220219 &        800 & 6.90 & 3.306 & 15.924 & $490.6 \pm 34.2$ \\
TIC 452522881 & Goodman/SOAR & 20220219 &        300 & 6.54 & 3.224 & 15.087 &            $< 10$\\ 
TIC 238714485 & Goodman/SOAR & 20220220 &        700 & 7.09 & 3.400 & 15.720 & $632.6 \pm 22.2$ \\ 
     HD 42270 &     NRES/LCO & 20220905 &       1500 & 3.10 & 0.990 &  8.915 & $229.9 \pm 11.3$ \\ 
     HD 21024 &     NRES/LCO & 20220822 &        600 & 2.14 & 0.585 &  5.406 &           $< 10$ \\
     HD 44627 &     NRES/LCO & 20220909 &       1200 & 3.48 & 1.082 &  8.838 & $183.2 \pm 13.6$ \\
    \hline
    \end{tabular}
    \label{table:observed}
\end{table*}

\subsection{Archival}\label{sec:carina_archival}
We obtain archival astrometry ($\alpha$, $\delta$, $\pi$, $\mu$), photometry ($B_P$, $R_P$, and $G$), radial velocities, and \gaia{} RUWE for association members from \gaia{} DR3 \citep{gaia_collaboration_gaia_2022}.

We also obtain the J, H, and $K_S$ magnitudes for all applicable members from the Two-Micron All-sky survey \citep[2MASS;][]{skrutskie_two_2006}.

\section{Revised age of Carina} \label{sec:age}
\subsection{Isochrone}

We first estimate the age of Carina by comparing \gaia{} photometry of stars in our updated membership list to two sets of isochrones: PARSECv1.2 \citep{bressan_parsec_2012} and the Dartmouth Stellar Evolution Program \citep[DSEP, ][]{dotter_dartmouth_2008} with magnetic enhancement \citep{feiden_magnetic_2016}. Both model grids have been shown to perform well on $10$--$150$ Myr associations like Carina \citep[e.g.,][]{gillen_new_2017, mann_tess_2022, wood_tess_2023}. 

For this comparison, we use a Gaussian mixture model following \citet{mann_tess_2022}. To briefly summarize, the likelihood is formed from the mixture of two models as described in \citet{hogg_data_2010}. The first model represents the single-star sequence of true members, and is described by two parameters: age ($\tau$) and reddening ($E(B-V)$). The second model captures the outliers, which may include binaries, non-members, or stars with problematic photometry or parallaxes. The outlier model is described with two parameters: the offset from the first model ($Y_B$) and the variance around that offset ($V_B$). There are two additional free parameters, one to describe the amplitude of the outlier model ($P_B$), and one to handle underestimated uncertainties in the model and/or photometry ($f$). More details, including the likelihood function, can be found in the appendix of \citet{mann_tess_2022}.

We apply some quality cuts to the membership list from Section~\ref{sec:membership}. Specifically, we remove stars with photometry outside the model grid, those with RUWE$>$1.4, and those with poor photometry or parallaxes (SNR of any absolute magnitude $<$20). 

\begin{figure}[tb]
    \centering
    \includegraphics[width=0.49\textwidth]{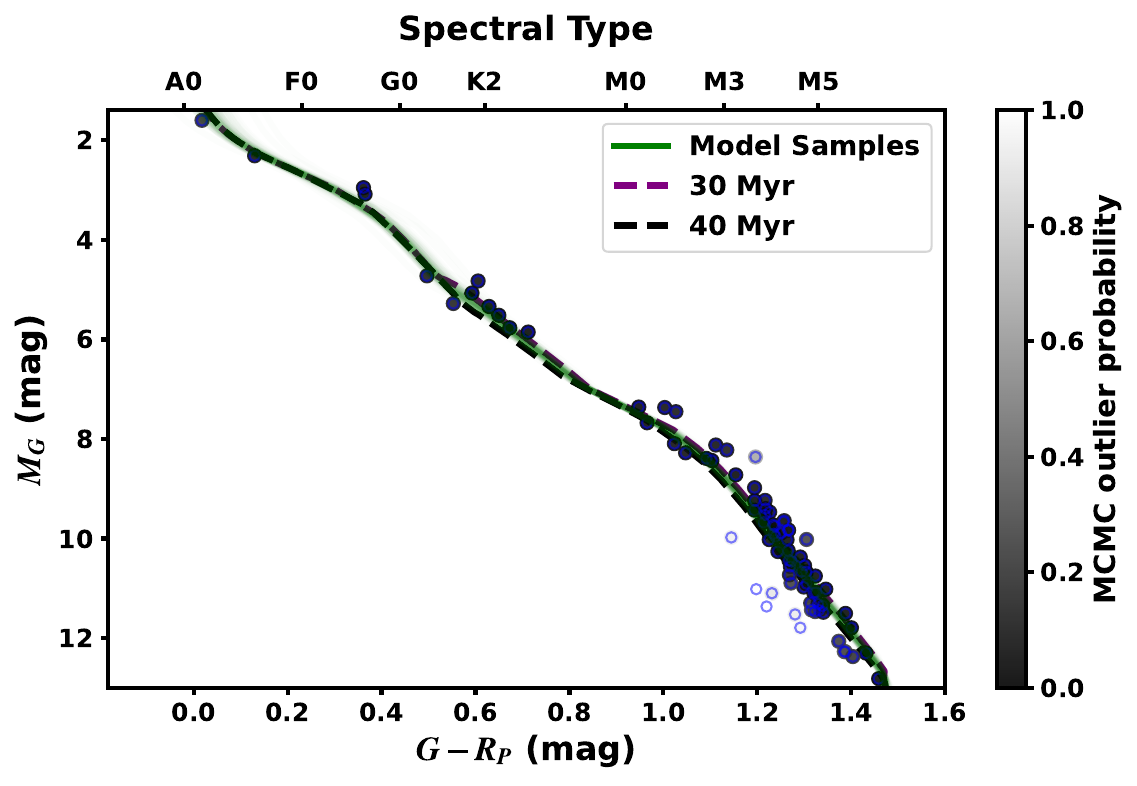}
    \caption{CMD of stars selected for fitting (blue circles) compared to PARSECv1.2 models. The green lines are 200 random samples from our MCMC analysis, and the two dashed lines are the 30- and 40-Myr isochrones for comparison. Points are shaded based on the probability that they are part of the main population or the outlier population. }
    \label{fig:isochrone}
\end{figure}

We wrap the likelihood in a Monte Carlo Markov Chain (MCMC) framework using \texttt{emcee} \citep{foreman-mackey_emcee_2013}. The MCMC is run for 10,000 steps after an initial burn-in of 1,000 steps, which far exceeds 50 times the autocorrelation time required for convergence. All parameters evolve under uniform priors with only physical limitations. Extinction is allowed to go negative to avoid Lucy-Sweeney bias \citep{lucy_significance_1979}; we expect such nearby stars to have minimal reddening. To ensure uniform sampling in age, we re-sample the underlying model isochrones to increments of 0.1 Myr using the \texttt{isochrones} package \citep{morton_isochrones_2015}. We assume Solar metallicity but tested near-Solar metallicities ($-0.3<$[M/H]$<0.3$). 
We show the \gaia{} CMD and model fit in Figure~\ref{fig:isochrone}. The final fit yields an age of $\tau=34\pm3$\,Myr when using the PARSEC models, and $\tau=39\pm3$\,Myr when using the DSEP magnetic models. Different input assumptions, such as locking $E(B-V)$ to zero, changes in metallicity, or changing the assumed solar abundance scale, shift the resulting age at the $\simeq$3\,Myr level, similar to the reported errors and the difference between the result of the two model grids. 

\subsection{Variability}\label{sec:var}
\citet{barber_using_2023} present a method to estimate the age of a group from the overall variability of the members. While less precise, it is based purely on the variation in the \gaia{} photometry, and hence is independent of the other two methods. Using the combination of all three \gaia{} bands yielded an age estimate of $29^{+13}_{-7}$\,Myr. This is within $1\sigma$ of our isochrone- and lithium-based estimates. 

\subsection{Lithium Depletion Boundary}\label{sec:ldb}

We use the EW(Li) measurements obtained above to determine the association age in two ways: the LDB method, and by comparing the full Li sequence to benchmark associations.

The LDB method relies upon locating the sharp cutoff between Li-rich and Li-poor M dwarfs within the association, caused by the rapid depletion of Li in the cores of fully-convective stars with core temperatures $> 2.5\times10^6 K$. As they approach the MS, stars of different masses will reach this threshold temperature at different times, resulting in a sharp boundary between those which have reached it and fully depleted their initial Li, and those slightly lower mass stars which have not.

\begin{figure}[tb]
    \centering
    \includegraphics[width=0.49\textwidth]{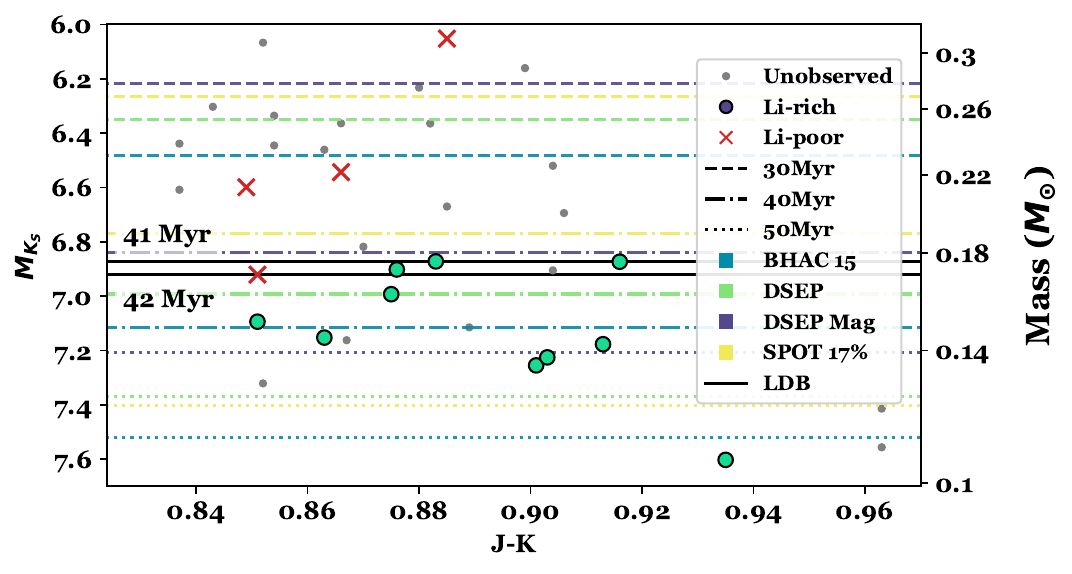}
    \caption{The Lithium Depletion Boundary of Carina. Li-rich stars (EW(Li) $> 200$m\AA) are  shown as teal dots, and Li-poor (EW(Li) $< 200$m\AA) ones as red x's. Black horizontal lines are drawn through the  brightest Li-rich star and the faintest Li-poor one, and labeled with the corresponding age using the DSEP magnetic model. The LDB at 30, 40, and 50 Myr for each model is shown as colored lines, where the color indicates the model, and the line style indicates the age. The right side axis shows the stellar mass corresponding to the given magnitude at 40 Myr.}
    \label{fig:ldb}
\end{figure}

Identifying this boundary requires defining the threshold between Li-rich and Li-poor stars \citep[see e.g.,][]{binks_lithium_2014, binks_gaia-eso_2021}. We use $EW(Li) = 200$m\AA\, as the threshold \citep[following][]{binks_lithium_2014}. Using this threshold, we find that 12 of the observed stars are Li-rich (see Table \ref{table:observed} for Li measurements). The edges of the LDB are defined by the faintest Li-poor and brightest Li-rich stars. We find that the LDB is $6.87 < M_{K_s} < 6.92$, shown in Figure \ref{fig:ldb}.

We determine the age of the association by comparing this magnitude range to stellar evolutionary models. We use six different models with varying treatments of convection, magnetic fields, and spots. For standard models we use the models from \citet[][BHAC15]{baraffe_new_2015} and \citet[][DSEP]{dotter_dartmouth_2008}. For models with treatment of magnetic fields we use the Dotter models with magnetic enhancement from \citet[][DSEP mag]{feiden_magnetic_2016}, and the stellar spot models from \citet{somers_spots_2020} with $17\%$ and $34\%$ spot coverage (SPOT). 
\red{The DSEP mag models are available based on stellar abundances from either \citet[][GS98]{grevesse_standard_1998} or \citet[][AGSS09]{asplund_chemical_2009}. We calculate the LDB age using both, but plot only the GS98 models on all figures.}

For each model we calculate the magnitude corresponding to $99\%$ Li depletion at each modeled age. We then linearly interpolate the resulting relationship between LDB magnitude and age to find the predicted age for the observed LDB. We repeat this interpolation using the top and bottom edges of the LDB to determine lower and upper age limits with each model, shown in Table \ref{table:ldb_ages}. To calculate an overall age estimate we take the average of all estimates, resulting in an age of $41.5 \pm 3.2$ Myr.

\begin{table}[ht!]
    \centering
    \begin{tabular}{|c|c|c|}
     \hline
     Model & Lower Bound & Upper Bound\\ 
     \hline
    BHAC15            & 36 Myr & 37 Myr \\
    DSEP  (GS98)      & 38 Myr & 39 Myr \\
    DSEP Mag  (GS98)  & 41 Myr & 42 Myr \\
    DSEP Mag (AGSS09) & 44 Myr & 45 Myr \\
    SPOT 17\%       & 42 Myr & 42 Myr \\
    SPOT 34\%       & 45 Myr & 47 Myr \\
     \hline   
    \end{tabular}
    \caption{Upper and lower age bounds given by each of the models used. The upper bound corresponds to the age given an LDB at the magnitude of the faintest observed Li-poor star, and the lower bound corresponds to the age given an LDB at the magnitude of the brightest observed Li-rich star.}
    \label{table:ldb_ages}
\end{table}

A higher threshold of $EW(Li) > 300 $ m\AA\, \citep[as in][]{binks_gaia-eso_2021} results in 3 of the Li-rich stars becoming Li-poor, for a total of 9 Li-rich stars. This changes the LDB bounds to be $6.87 < M_{K_s} < 7.25$, resulting in a slightly older age of $45.3 \pm 5.7$ Myr, which is still within the $2\sigma$ uncertainty bounds.

Another source of uncertainty in the LDB age measurement are the unobserved association members. 
There are several known members with magnitudes between the upper edge of the LDB and the next observed star. If some or all of these 4 stars are Li-rich, the top edge of the LDB would be shifted to a brighter magnitude, resulting in a younger age. 
The brightest unobserved star in this magnitude range is at $M_{K_s} = 6.608$. Using this as the lower age boundary results in an age of $38\pm4$ Myr, within the uncertainty bounds of our measurement.

\subsection{Lithium Sequence}\label{sec:ew_li}
The Li sequence is an alternative method which utilizes Li measurements across a larger range of stellar masses. By comparing Li abundance as a function of color to either evolutionary models or benchmark associations, an age can be determined \citep{soderblom_ages_2014}. When comparing to evolutionary models the dependence on initial Li abundance and greater reliance on factors such as convective overshoot make this method less robust to changes in model and less consistent than LDB. Using empirical comparison, the resulting age is only as model dependent as the age measurements of the benchmark association it is based on. \red{Comparison of the Li sequence to models in particular often produces ages which are younger than those derived using the LDB.}

To construct the Li sequence of Carina we use our Li measurements of low- and moderate-mass stars (see Table \ref{table:observed}), supplemented with additional measurements from \citet{riedel_lacewing_2017} and \citet{schneider_acronym_2019}.  We obtain $13$ measurements from \citet{riedel_lacewing_2017}, and $3$ from \citet{schneider_acronym_2019}, listed in Table \ref{table:carina_lit_li}. Of the $13$ in \citet{riedel_lacewing_2017}, they classify eight as Carina members, while the remaining are either members of other nearby young associations (TW Hydrae and Tuc-Hor) or were not assigned membership in any association. We find that all have $P_{BANYAN} > 60\%$.

\red{To compare the observed stars against evolutionary models we first converted the measured EW(Li) to fraction of initial Li remaining, Li/Li$_0$. To do so we determined initial Li abundance for each star using a curve of growth. The curve of growth for stars with $\teff{} < 4000$K was taken from \citet{zapatero_osorio_lithium_2002} and that for stars with $\teff > 4000$K from \citet{soderblom_stellar_1993}. We then divided the EW(Li) by the intial EW(Li) to produce the Li fraction.}

\begin{figure*}
    \centering
    \includegraphics[width=0.49\linewidth]{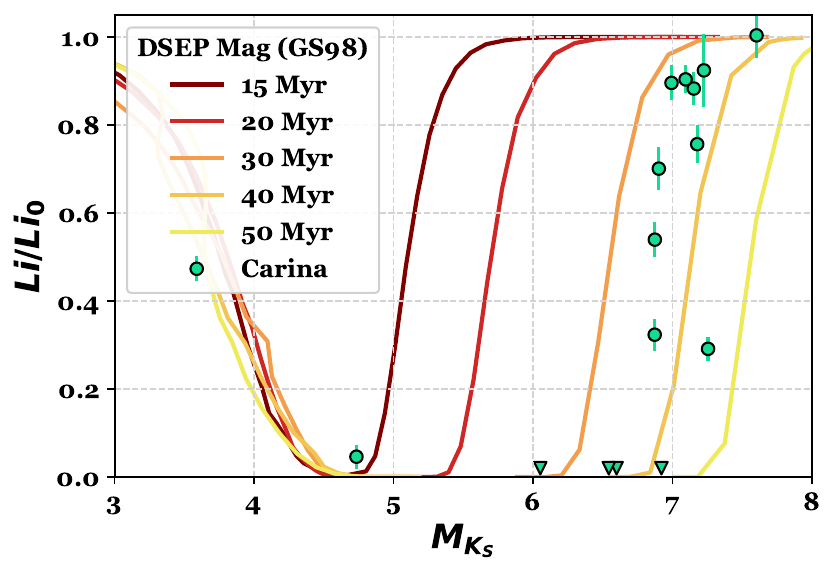} 
    \includegraphics[width=0.49\linewidth]{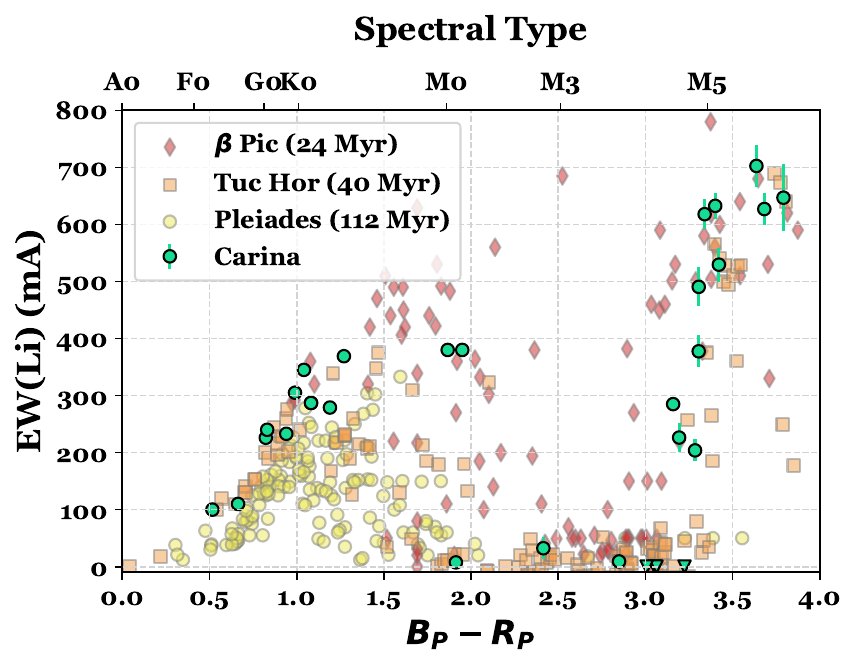}
    \caption{Lithium measurements of Carina members.
    Left)  Fraction of initial Li abundance as a function of \gaia{} $B_P-R_P$, overplotted with the DSEP magnetic models \citep[][DSEP mag]{feiden_magnetic_2016}. Equivalent widths of association members were converted to Li/Li$_0$ using the curve of growth from \cite{zapatero_osorio_lithium_2002} for $\teff{} < 4000 K$ and from \citet{soderblom_stellar_1993} for $\teff > 4000 K$.
    Right) Lithium equivalent widths for Carina members, from our observations and \citet{riedel_lacewing_2017, schneider_acronym_2019}, are shown as teal dots plotted on top of the observed Li Sequences of the Beta Pictoris Moving Group \citep[$24$ Myr, red;][]{binks_lithium_2014}, Tucana-Horologium Young Association \citep[$40$ Myr, orange;][]{kraus_stellar_2014}, and the Pleiades \citep[$112$ Myr, yellow;][]{bouvier_lithium-rotation_2018}.}
    \label{fig:carina_li_seq}
\end{figure*}

In Figure \ref{fig:carina_li_seq} we show both the empirical comparison of the Carina member Li measurements against benchmark associations and a comparison of the Li fraction to stellar evolutionary models.

\red{Empirically, the Li sequence of Carina lies very near that of the Tuc-Hor association, shown in the right panel of Figure \ref{fig:carina_li_seq}. The association has less Li at $B_P-R_P \simeq 3.0$ than the $\beta$ Pic moving group ($24$ Myr), and does not transition to Li-rich stars until $B_P-R_P \simeq 3.25$. \citet{kraus_stellar_2014} found that Tuc-Hor has a LDB at $M_{K_s}=7.12\pm0.16$. In that work they calculated an age of $41\pm2$ using that LDB and models from \citet{baraffe_evolutionary_1998}. Using the method described in Section \ref{sec:ldb} with $M_{K_s}=7.12\pm0.16$, we find an age for Tuc-Hor of $45.7\pm4.7$ Myr. This makes Tucana-Horologium slightly older than Carina, indicating that the empirical Li sequence of Carina is consistent with our calculated LDB age.}

\red{Comparing Li fraction against models shows that most of the points fall between the $30$ Myr and $40$ Myr isochrones. This is somewhat younger than the previous estimates, but still consistent within the uncertainties.}

\begin{table*}
    \centering
    \caption{Literature Li measurements for Carina members.}
    \vspace{5pt}
    \begin{tabular}{|lcccccc|}
    \hline
Object &  Gaia DR3 & $M_{K_s}$ & $B_P-R_P$ & $G$ & EW(Li) &  Source \\
    &   & mag & mag & mag & mA & \\ 
\hline
HD 49855    & 5265670762922792960 & 3.561 & 0.940 & 8.995 & 233.0 & \citet{riedel_lacewing_2017} \\
TWA 21      & 5356713413789909632 & 3.569 & 1.271 & 9.477 & 369.0 & \citet{riedel_lacewing_2017} \\
HD 42270    & 4621305817457618176 & 3.096 & 0.990 & 8.915 & 305.0 & \citet{riedel_lacewing_2017} \\
AB Pic     & 5495052596695570816 & 3.48  & 1.082 & 8.838 & 287.0 & \citet{riedel_lacewing_2017} \\
HD 37402    & 4759444786175885824 & 2.828 & 0.664 & 8.267 & 110.0 & \citet{riedel_lacewing_2017} \\
2MASS J04082685-7844471 & 4625883599760005760 & 4.597 & 1.913 & 11.48  & 7.5 & \citet{riedel_lacewing_2017} \\
HD 55279    & 5208216951043609216 & 3.64  & 1.191 & 9.777 & 279.0 & \citet{riedel_lacewing_2017} \\
V0479 Car   & 5299141546145254528 & 3.043 & 1.041 & 9.747 & 345.0 & \citet{riedel_lacewing_2017} \\
2MASS J02564708-6343027 & 4721078629298085760 & 5.165 & 2.849 & 12.803 & 9.2 & \citet{riedel_lacewing_2017} \\
HD 269920   & 4658442922197295232 & 3.314 & 0.823 & 9.471 & 226.0 & \citet{riedel_lacewing_2017} \\
HD 83096*    & 5217846851839896704 & 1.71  & 0.516 & 7.413 & 100.0 & \citet{riedel_lacewing_2017} \\
HD 83096 B*  & \red{5217846851839896832} & \red{-} & \red{0.804} & \red{9.193} & 240.0 & \citet{riedel_lacewing_2017} \\
2MASS J07065772-5353463 & 5491506843495850240 & 4.314 & 1.866 & 10.71  & 380.0 & \citet{schneider_acronym_2019} \\
2MASS J09032434-6348330 & 5297100607744079872 & 4.228 & 1.949 & 11.912 & 380.0 & \citet{schneider_acronym_2019} \\
2MASS J09180165-5452332 & 5310606291358320512 & 5.127 & 3.159 & 13.15  & 285.0 & \citet{schneider_acronym_2019} \\
     \hline   
    \end{tabular}
    \tablecomments{\red{*Binary, unresolved in 2MASS, but resolved in GaiaDR3.}}
    \label{table:carina_lit_li}
\end{table*}

\section{Summary and Discussion}\label{sec:discussion}

We measured the age of the Carina association using new Li measurements and the LDB method. Using \gaia{} DR3 kinematic measurements, and \banyan{} code with an updated list of associations, we created a new membership list of Carina, containing $99$ stars. We obtained medium-resolution optical spectra  of $15$ low-mass association members using the Goodman HTS on the 4m SOAR telescope, and supplemented these with spectra of K-type members taken with NRES on LCO. From these spectra we measured EW(Li), located the LDB, and constructed a Li sequence of the association. 
\red{We supplemented the Li-based age measurements with a Gaussian-mixture model CMD fit, and analysis of the \gaia{} photometric variability.}
\red{By combining all of the age measurements we obtain an age for the association of $41^{+3}_{-5}$ Myr. This includes estimates based on CMD position, photometric variability, LDB, and comparison of the Li sequence to benchmark associations and models.}

Our age measurement is consistent with \cite{bell_self-consistent_2015}.
However, it is much older than the age found by \cite{booth_age_2021}. That age was largely based on the inclusion of a single high-mass star, HD 95086, in the membership of Carina. \citet{wood_tess_2023} found that that star is instead a high-probability member of the newly-discovered MELANGE-4 association.

Our lithium-based age is also $\sim 2\times$ the age found by \citet{schneider_acronym_2019} using a compilation of Li measurements. They found that the Li sequence of Carina more closely resembled the Li sequence of Beta Pic ($\sim 21$ Myr) than that of Tuc-Hor ($\sim40$ Myr). Differentiation between those two ages comes largely from three stars with spectral type M0 and $EW(Li) = 350 - 400 $m\AA, which is higher than expected for a $40$ Myr old star of that spectral type. Two of those three stars are included in the membership list we use here.
Our age is based on Li measurements in fully-convective low-mass stars, as there is less scatter in the Li levels for a single-aged populations in these stars than in warmer, partially radiative FGK and early M dwarfs. New Li measurements in stars of type M3 and later make clear that the LDB for Carina, shown in Figures \ref{fig:ldb} and \ref{fig:carina_li_seq}, is at a higher magnitude than that for $\beta$ Pic. If Carina were the same age as $\beta$ Pic, we would expect the three observed stars with $6.0 < M_{K_S} < 6.5 $ to be Li-rich, and the partially depleted stars at $6.5 < M_{K_S} < 7.0 $ not to be depleted at all. 
This discrepancy emphasizes the utility of the LDB method, which is less sensitive to model selection and which operates in a stellar regime with less star-to-star variation in lithium levels.

\red{While we report a single age for Carina, not all members of an association form at the same time, leading to age spread within an association. However, this spread is difficult to disentangle from other factors, as any variation in an age-dependent property between members of an association may be caused by true age spread or by measurement error or age-unrelated variation in the property. Differences in stellar accretion history \citep{baraffe_effect_2010, baraffe_self-consistent_2017}, magnetic activity and spot fraction \citep[e.g.][]{binks_gaia-eso_2021}, extinction, circumstellar disks, binarity \citep{sullivan_undetected_2021}, and rotation may all appear as an age spread. 
It is possible that large age spreads are more common in stellar association complexes, such as Scorpius-Centaurus, where multiple formation events may have occurred over a $\sim10$ Myr period. Age spreads of $1-6$ Myr have been measured in the Taurus complex \citep{krolikowski_gaia_2021}, and of $6-7$ Myr within populations of Scorpius-Centaurus \citep{pecaut_star_2016}. So an upper limit estimate of potential age spread within Carina would be $\pm6$ Myr.}

\red{While the LDB and Li sequence of the observed Carina members is very tight, there is one star with a lower than expected EW(Li) compared to its $M_{K_s}$ (TIC 350559457). If the low Li is caused by age spread alone, based on the Li/Li$_0$ vs. $M_{K_s}$ sequence, this star would indicate age spread of 10 Myr (see Figure \ref{fig:carina_li_seq}).
However, its EW(Li) is very similar to other association members with a similar $B_P-R_P$ color. 
Additionally, none of the other observed stars differ by much from the expected Li for the determined age. The other methods we use are similarly consistent, with low CMD spread and a narrow LDB.
Thus, it seems more likely that the low Li sequence position of TIC 350559457 is caused, at least in part, by measurement error or a different astrophysical reason than for it to be entirely from age spread.}

The age we find here is consistent with the age of the nearby Tuc-Hor association \citep{kraus_stellar_2014}, lending support to the theory that the three groups form a complex as suggested early after their discovery \citep{torres_great_2001}  and by recent work \citep[e.g.,][]{kerr_stars_2021}. Other potentially related groups include Theia 92, Theia 113 and Platais 8 \citep{gagne_number_2021}. Measuring LDB ages for those other groups to confirm their ages is an important next step to mapping out the larger star-forming complex.

If these groups, or a subset of them, are related, then this structure could be an older remnant of a Sco-Cen-like complex. Historically, older regions have been much harder to study because associations spread out as they age and galactic forces pull them apart. This has prevented the identification and study of older complexes --- the only two well-studied complexes are both less than $20$ Myr old. 
If so, this complex can reveal new insight into star formation and molecular cloud collapse mechanisms as well as to the stages of stellar and planetary evolution between $20$--$100$ Myr.

\newpage{}
\begin{acknowledgments}

A.W.M., M.G.B., and M.L.W were supported by an NSF CAREER grant (AST-2143763). M.L.W., P.C.T., and R.P.M. were supported by the NC Space Grant Graduate Research program. P.C.T. was also supported by NSF Graduate Research Fellowship (DGE-1650116), the Zonta International Amelia Earhart Fellowship, and the Jack Kent Cooke Foundation Graduate Scholarship. 

Many thanks to Patricio, Carlos, Juan, Sergio, and Rodrigo at SOAR for helping through many nights of observations. 

This research includes data from observations obtained at the Southern Astrophysical Research (SOAR) telescope, which is a joint project of the Minist\'{e}rio da Ci\^{e}ncia, Tecnologia e Inova\c{c}\~{o}es (MCTI/LNA) do Brasil, the US National Science Foundation’s NOIRLab, the University of North Carolina at Chapel Hill (UNC), and Michigan State University (MSU).

This work has made use of data from the European Space Agency (ESA) mission \textit{Gaia} (\url{https://www.cosmos.esa.int/gaia}), processed by the \textit{Gaia} Data Processing and Analysis Consortium (DPAC) (\url{https://www.cosmos.esa.int/web/gaia/dpac/consortium}). Funding for the DPAC has been provided by national institutions, in particular the institutions participating in the \emph{Gaia} Multilateral Agreement.  

This research has made use of the VizieR catalogue access tool, CDS, Strasbourg, France. The original description of the VizieR service was published in A\&AS 143, 23. Resources supporting this work were provided by the NASA High-End Computing (HEC) Program through the NASA Advanced Supercomputing (NAS) Division at Ames Research Center for the production of the SPOC data products. 

This work makes use of observations from the LCOGT NRES network.

\end{acknowledgments}

%

\vspace{5mm}
\facilities{SOAR 4m (Goodman HTS), LCOGT 1m (NRES)}


\software{
Astropy \citep{astropy_collaboration_astropy_2013, astropy_collaboration_astropy_2018},
Astroquery \citep{ginsburg_astroquery_2019},
matplotlib \citep{hunter_matplotlib_2007},
\banyan{} \citep{malo_bayesian_2012, gagne_banyan_2018},
BANZAI-NRES,
}

\bibliography{biblio}{}
\bibliographystyle{aasjournal}

\end{document}